\documentclass[twocolumn,preprintnumbers,amsmath,amssymb]{revtex4}
\usepackage{graphicx}
\usepackage{amssymb}

\begin{document}

\title{Copenhagen interpretation can survive the upgraded Schr\"{o}dinger's
cat Gedankenexperiment}
\author{Guang Ping He}
\email{hegp@mail.sysu.edu.cn}
\affiliation{School of Physics, Sun Yat-sen University, Guangzhou 510275, China}

\begin{abstract}
Recently, Frauchiger and Renner proposed a Gedankenexperiment, which was
claimed to be able to prove that quantum theory cannot consistently describe
the use of itself. Here we show that the conclusions of Frauchiger and
Renner actually came from their incorrect description of some quantum
states. With the correct description there will be no inconsistent results,
no matter which quantum interpretation theory is used. Especially, the
Copenhagen interpretation can satisfy all the three assumptions (C), (Q),
and (S) of Frauchiger and Renner simultaneously, thus it has no problem
consistently describing the use of itself.
\end{abstract}


\maketitle


\section{Introduction}

In a recent publication of Frauchiger and Renner \cite{qi1571}, a
Gedankenexperiment was proposed, which was claimed to be able to lead to
inconsistent conclusions with a self-referential use of quantum theory. Thus
it seems to prove that quantum theory cannot consistently describe the use
of itself. The Gedankenexperiment was highly praised to be \textquotedblleft
more involved than Schr\"{o}dinger's cat\textquotedblright\ \cite%
{news,qi1577}, and caught great interests immediately \cite%
{qi1631,qi1572,qi1575,qi1592,qi1608,qi1680,qi1683}. Here we show that the
original conclusions of Frauchiger and Renner actually came from their own
inconsistent use of quantum theories in the description of the quantum systems. With the correct description
there will be no inconsistent results. Especially, the Copenhagen
interpretation can satisfy all the three assumptions (C), (Q), and (S) of
Frauchiger and Renner simultaneously, thus it has no problem consistently
describing the use of itself. Unlike \cite%
{qi1631,qi1572,qi1575,qi1592,qi1608,qi1680,qi1683}, our result reveals that
there is no need to introduce new hypotheses in quantum theory or modify the
original Gedankenexperiment.

The next section is a brief review on the Gedankenexperiment and the result
of Frauchiger and Renner. Then in Sect. III, we will elaborate what went
wrong in Frauchiger and Renner original reasoning, and why the inconsistency
will disappear when applying quantum interpretation theories correctly. In
Sect. IV we will address the lesson that we learned from this case. Finally, our conclusions will be summarized and compared with related works from other researchers in Sect. V.

\section{The Frauchiger-Renner Gedankenexperiment}

In brief, Frauchiger and Renner proposed the following three assumptions in
Boxes 2-4 of \cite{qi1571}.

\bigskip

\textit{Assumption (Q)}

Suppose that agent $A$ has established that

\qquad \textit{Statement} $A^{(i)}$: \textquotedblleft System $S$ is in
state $\left\vert \psi \right\rangle _{S}$\ at time $t_{0}$%
.\textquotedblright

Suppose furthermore that agent $A$ knows that

\qquad \textit{Statement} $A^{(ii)}$: \textquotedblleft The value $x$ is
obtained by a measurement of $S$ w.r.t. the family $\{\pi
_{x}^{t_{0}}\}_{x\in \chi }$\ of Heisenberg operators relative to time $%
t_{0} $, which is completed at time $t$.\textquotedblright

If $\left\langle \psi \right\vert \pi _{\xi }^{t_{0}}\left\vert \psi
\right\rangle =1$\ for some $\xi \in \chi $\ then agent $A$ can conclude that

\qquad \textit{Statement} $A^{(iii)}$: \textquotedblleft I am certain that $%
x=\xi $ at time $t$.\textquotedblright

\bigskip

\textit{Assumption (C)}

Suppose that agent $A$ has established that

\qquad \textit{Statement} $A^{(i)}$: \textquotedblleft I am certain that
agent $A^{\prime }$, upon reasoning within the same theory as the one I am
using, is certain that $x=\xi $ at time $t$.\textquotedblright

Then agent $A$ can conclude that

\qquad \textit{Statement} $A^{(ii)}$: \textquotedblleft I am certain that $%
x=\xi $ at time $t$.\textquotedblright

\bigskip

\textit{Assumption (S)}

Suppose that agent $A$ has established that

\qquad \textit{Statement} $A^{(i)}$: \textquotedblleft I am certain that $%
x=\xi $ at time $t$.\textquotedblright

Then agent $A$ must necessarily deny that

\qquad \textit{Statement} $A^{(ii)}$: \textquotedblleft I am certain that $%
x\neq \xi $ at time $t$.\textquotedblright

\bigskip

Then the main conclusion of \cite{qi1571} is the following no-go theorem.

\bigskip

\textit{Theorem 1}

Any theory that satisfies assumptions (Q), (C), and (S) yields contradictory
statements when applied to the Gedankenexperiment of Box 1 (of \cite{qi1571}%
).

\bigskip

This Gedankenexperiment is a procedure among four agents $\bar{F}$, $F$, $%
\bar{W}$\ and $W$. They repeat the following steps in rounds $n=0,1,2,...$
until the halting condition in the last step is satisfied:

At time $n:00$, Agent $\bar{F}$\ measures a quantum system $R$ in state%
\begin{equation}
\left\vert init\right\rangle _{R}=\sqrt{\frac{1}{3}}\left\vert
heads\right\rangle _{R}+\sqrt{\frac{2}{3}}\left\vert tails\right\rangle _{R}
\label{init}
\end{equation}%
and denote the outputs as $r=heads$ or $r=tails$. She sets the spin $S$ of a
particle to $\left\vert \downarrow \right\rangle _{S}$ if $r=heads$, and to $%
\left\vert \rightarrow \right\rangle _{S}\equiv \sqrt{1/2}(\left\vert
\downarrow \right\rangle _{S}+\left\vert \uparrow \right\rangle _{S})$ if $%
r=tails$, and sends it to $F$.

At $n:10$, Agent $F$ measures $S$ in the basis $\{\left\vert \downarrow
\right\rangle _{S},\left\vert \uparrow \right\rangle _{S}\}$, recording the
outcome $z\in \{-1/2,+1/2\}$.

At $n:20$, Agent $\bar{W}$ measures lab $\bar{L}$ (containing system $R$ and
agent $\bar{F}$) in a basis containing%
\begin{equation}
\left\vert \overline{ok}\right\rangle _{\bar{L}}=\sqrt{\frac{1}{2}}\left(
\left\vert \bar{h}\right\rangle _{\bar{L}}-\left\vert \bar{t}\right\rangle _{%
\bar{L}}\right) .  \label{okL}
\end{equation}%
Here $\left\vert \bar{h}\right\rangle _{\bar{L}}$ and $\left\vert \bar{t}%
\right\rangle _{\bar{L}}$ are defined as the states of lab $\bar{L}$\ at the
end of the first step, depending on whether $r=heads$\ or $r=tails$,
respectively. If the outcome associated to $\left\vert \overline{ok}%
\right\rangle _{\bar{L}}$ occurs he announces $\bar{w}=\overline{ok}$\ and
else $\bar{w}=\overline{fail}$.

At $n:30$, Agent $W$ measures lab $L$ (containing system $S$ and agent $F$)
in a basis containing%
\begin{equation}
\left\vert ok\right\rangle _{L}=\sqrt{\frac{1}{2}}\left( \left\vert -\frac{1%
}{2}\right\rangle _{L}-\left\vert +\frac{1}{2}\right\rangle _{L}\right) .
\end{equation}%
Here $\left\vert -1/2\right\rangle _{L}$ and $\left\vert +1/2\right\rangle
_{L}$\ are defined as the state of lab $L$\ depending on whether the
incoming spin was $\left\vert \downarrow \right\rangle _{S}$ or $\left\vert
\uparrow \right\rangle _{S}$, respectively. If the outcome associated to $%
\left\vert ok\right\rangle _{L}$ occurs he announces $w=ok$\ and else $%
w=fail $.

At $n:40$, if $\bar{w}=\overline{ok}$\ and $w=ok$\ then the experiment is
halted.

\bigskip

According to \cite{qi1571}, this Gedankenexperiment can make the agents
arrive at conflicting conclusions (thus violating Assumption (C)) when they
all employ the same form of quantum theory. Some main points of the
reasoning is briefly reviewed below.

(i) From agent $\bar{F}$'s point of view:

Suppose that $\bar{F}$\ got $r=tails$\ in round $n$. Then according to the
above experimental instructions, she can make the statement

\textit{Statement} $\bar{F}^{n:01}$: \textquotedblleft The spin $S$ is in
state $\left\vert \rightarrow \right\rangle _{S}$\ at time $n:10$%
.\textquotedblright

Agent $\bar{F}$\ could then conclude that the later state of lab $L$ is%
\begin{equation}
U_{S\rightarrow L}^{10\rightarrow 20}\left\vert \rightarrow \right\rangle
_{S}=\sqrt{\frac{1}{2}}\left( \left\vert -\frac{1}{2}\right\rangle
_{L}+\left\vert +\frac{1}{2}\right\rangle _{L}\right)  \label{not ok}
\end{equation}%
(as presented in the paragraph before Eq. (4) of \cite{qi1571}), which is
orthogonal to $\left\vert ok\right\rangle _{L}$. Thus $\bar{F}$ will further
infer that

\textit{Statement} $\bar{F}^{n:02}$: \textquotedblleft I am certain that $W$
will observe $w=fail$ at time $n:31$.\textquotedblright

(ii) From agent $F$'s point of view:

Suppose that $F$ observed $z=+1/2$\ in this round. Note that if $\bar{F}$\
got $r=heads$ at time $n:01$, she should have sent $F$ the state $\left\vert
\downarrow \right\rangle _{S}$,\ which can never be observed as $z=+1/2$.
Therefore, $F$ can conclude that

\textit{Statement} $F^{n:12}$: \textquotedblleft I am certain that $\bar{F}$
knows that $r=tails$ at time $n:01$.\textquotedblright

Then $F$ can further infer from \textit{Statement} $\bar{F}^{n:02}$\ that

\textit{Statement} $F^{n:13}$: \textquotedblleft I am certain that $\bar{F}$
is certain that $W$ will observe $w=fail$ at time $n:31$.\textquotedblright

This is because $F$ infers that when $\bar{F}$ got $r=tails$, the state of
lab $L$ is Eq. (\ref{not ok}) which can never lead to the outcome $w=ok$.

(iii) From agent $\bar{W}$'s point of view:

Agent $\bar{W}$\ can be certain\ that there will always be $(\bar{w},z)\neq (%
\overline{ok},-1/2)$, i.e., the outcomes $\bar{w}=\overline{ok}$\ and $%
z=-1/2 $\ can never occur simultaneously. This is because, according to \cite%
{qi1571}, the state of $\bar{L}\otimes S$ at time $n:10$\ is%
\begin{equation}
U_{R\rightarrow \bar{L}S}^{00\rightarrow 10}\left\vert init\right\rangle
_{R}=\sqrt{\frac{1}{3}}\left\vert \bar{h}\right\rangle _{\bar{L}}\otimes
\left\vert \downarrow \right\rangle _{S}+\sqrt{\frac{2}{3}}\left\vert \bar{t}%
\right\rangle _{\bar{L}}\otimes \left\vert \rightarrow \right\rangle _{S}
\label{pure}
\end{equation}%
(as presented in the paragraph before Eq. (6) of \cite{qi1571}), which is
orthogonal to $\left\vert \overline{ok}\right\rangle _{\bar{L}}\otimes
\left\vert \downarrow \right\rangle _{S}$. Consequently, in any round once $%
\bar{W}$\ obtains $\bar{w}=\overline{ok}$, he can be sure that

\textit{Statement} $\bar{W}^{n:22}$: \textquotedblleft I am certain that $F$
knows that $z=+1/2$ at time $n:11$.\textquotedblright

Combining with the reasoning in the above point (ii), he can infer from
\textit{Statements} $F^{n:12}$, $\bar{F}^{n:02}$ and $F^{n:13}$ that

\textit{Statement} $\bar{W}^{n:23}$: \textquotedblleft I am certain that $F$
is certain that $W$ will observe $w=fail$ at time $n:31$.\textquotedblright

Putting points (i), (ii) and (iii) together, we see that once $\bar{W}$\
obtains $\bar{w}=\overline{ok}$, there should always be $w=fail$. Thus the
halting condition (i.e., both $\bar{w}=\overline{ok}$\ and $w=ok$\ occur) in
the last step of the Gedankenexperiment can never be satisfied.

But according to Eq. (7) of \cite{qi1571}, $\bar{w}=\overline{ok}$\ and
$w=ok$\ can occur in the same round with probability $1/12$. Therefore, when
this case indeed occurs, a contradiction is reached. (See page 4 and Table 3
of \cite{qi1571} for full details.)

\section{Our result}

In our opinion, however, the above contradiction (i.e., the violation of
Assumption (C)) actually comes from the mistaken reasoning in \cite{qi1571},
not from quantum theory itself. There are two different ways to see where
the mistake lies, depending on which quantum interpretation theory is used.

\bigskip

\textit{(1) In quantum interpretation theories which believe that
wavefunctions collapse when being measured.}

A prominent example is the Copenhagen interpretation. While it has many
variations, according to one widely accepted version, \textquotedblleft
central to the popular image of the Copenhagen interpretation is the idea
that observation-induced wave packet collapse is a mode of dynamical
evolution unique to measurement interactions\textquotedblright\ (as stated
at the beginning of Sect. 2 of \cite{qi1654}, see also \cite%
{Rosenfeld,Stapp,qi1691}). From this point of view, the above Eqs. (\ref{not
ok}) and (\ref{pure}) that Ref. \cite{qi1571} introduced (as appeared in the
paragraphs before Eqs. (4) and (6) of \cite{qi1571}, respectively) are not
the correct description of the corresponding states.

To make it more obvious, let us write the initial state of quantum system $R$
that $\bar{F}$\ will measure at time $n:00$ as%
\begin{equation}
\left\vert init\right\rangle _{R}=\sqrt{\frac{1}{3}}\left\vert
heads\right\rangle _{R}+e^{i\theta }\sqrt{\frac{2}{3}}\left\vert
tails\right\rangle _{R}.
\end{equation}%
When taking the phase difference $\theta =0$\ it is exactly the same as Eq. (%
\ref{init}) that appeared in the Gedankenexperiment in \cite{qi1571}. From
the Copenhagen interpretation point of view, \textquotedblleft every act of
observation is by its very nature an \textit{irreversible}
process\textquotedblright , \textquotedblleft the correlations between ...
different states disappear when the state of equilibrium is reached; they
are wiped out ... in the process of recording the result of the
measurement\textquotedblright\ (see the 2nd paragraph of page 7 of \cite%
{qi1658}). Therefore, after $\bar{F}$\ measures $\left\vert
init\right\rangle _{R}$ at time $n:00$ in the basis $\{\left\vert
heads\right\rangle _{R},\left\vert tails\right\rangle _{R}\}$, the
Copenhagen interpretation implies that $\left\vert init\right\rangle _{R} $\
collapses to either $\left\vert heads\right\rangle _{R}$ or $\left\vert
tails\right\rangle _{R}$ \textit{irreversibly}. Either way, the information
on $\theta $\ is lost. Especially, even there was $\theta \neq 0$\ at the
beginning, after the measurement the resultant state is still $\left\vert
heads\right\rangle _{R}$ or $\left\vert tails\right\rangle _{R}$, which
shows no difference from the $\theta =0$\ case. Thus, recovering the
original value of $\theta $ from the resultant state is impossible.
Consequently, there is no reason to believe that the later states of $\bar{L}%
\otimes S$ should take the form $U_{R\rightarrow \bar{L}S}^{00\rightarrow
10}\left\vert init\right\rangle _{R}=\sqrt{\frac{1}{3}}\left\vert \bar{h}%
\right\rangle _{\bar{L}}\otimes \left\vert \downarrow \right\rangle
_{S}+e^{i\theta }\sqrt{\frac{2}{3}}\left\vert \bar{t}\right\rangle _{\bar{L}%
}\otimes \left\vert \rightarrow \right\rangle _{S}$ with $\theta $ preserved
in its original value, because this is a \textit{reversible} unitary
evolvement of $\bar{L}\otimes S$, not a measurement in the sense of the
Copenhagen interpretation. Instead, after the measurement there should no
longer be any fixed phase difference between $\left\vert \bar{h}%
\right\rangle _{\bar{L}}\otimes \left\vert \downarrow \right\rangle _{S}$
and $\left\vert \bar{t}\right\rangle _{\bar{L}}\otimes \left\vert
\rightarrow \right\rangle _{S}$.

Therefore, after $\bar{F}$\ measured system $R$, from the outcome she
herself knows that $\bar{L}\otimes S$ will be in a pure state $\left\vert
\bar{h}\right\rangle _{\bar{L}}\otimes \left\vert \downarrow \right\rangle
_{S}$\ or $\left\vert \bar{t}\right\rangle _{\bar{L}}\otimes \left\vert
\rightarrow \right\rangle _{S}$\ with probability $1/3$ or $2/3$,
respectively. But from agent $\bar{W}$'s point of view, as he has no
information on $\bar{F}$'s measurement outcome before he makes his own
measurement, he does not know the purification of the state of $\bar{L}%
\otimes S$. Then the correct description of the state of $\bar{L}\otimes S$
is no longer Eq. (\ref{pure}) (which is a pure state with a fixed phase
difference $\theta =0$). Instead, $\bar{L}\otimes S$ will appear to $\bar{W}$
as a mixture (which has no fixed phase difference between its components)
described by the density matrix%
\begin{eqnarray}
\rho _{\bar{L}S}^{10} &=&\frac{1}{3}\left( \left\vert \bar{h}\right\rangle _{%
\bar{L}}\otimes \left\vert \downarrow \right\rangle _{S}\right) \left(
\left\langle \downarrow \right\vert _{S}\otimes \left\langle \bar{h}%
\right\vert _{\bar{L}}\right)  \nonumber \\
&&+\frac{2}{3}\left( \left\vert \bar{t}\right\rangle _{\bar{L}}\otimes
\left\vert \rightarrow \right\rangle _{S}\right) \left( \left\langle
\rightarrow \right\vert _{S}\otimes \left\langle \bar{t}\right\vert _{\bar{L}%
}\right) .  \label{mix}
\end{eqnarray}

For the same reason, after $F$ measured $\left\vert \rightarrow
\right\rangle _{S}$ in the basis $\{\left\vert \downarrow \right\rangle
_{S},\left\vert \uparrow \right\rangle _{S}\}$ at time $n:10$, the
wavefunction collapsed from $\left\vert \rightarrow \right\rangle _{S}$ to
either $\left\vert \downarrow \right\rangle _{S}$ or $\left\vert \uparrow
\right\rangle _{S}$, and the information on the phase difference between $%
\left\vert \downarrow \right\rangle _{S}$ and $\left\vert \uparrow
\right\rangle _{S}$\ in $\left\vert \rightarrow \right\rangle _{S}=\sqrt{1/2}%
(\left\vert \downarrow \right\rangle _{S}+\left\vert \uparrow \right\rangle
_{S})$ is lost. Thus, from agent $W$'s point of view, the correct
description of the states of lab $L$ is no longer the pure state shown in
Eq. (\ref{not ok}). Instead, lab $L$ will be in a mixture described by the
density matrix%
\begin{equation}
\rho _{L}^{20}=\frac{1}{2}\left( \left\vert -\frac{1}{2}\right\rangle
_{L}\left\langle -\frac{1}{2}\right\vert +\left\vert +\frac{1}{2}%
\right\rangle _{L}\left\langle +\frac{1}{2}\right\vert \right) .  \label{ok}
\end{equation}

With these correct descriptions, we can see that unlike Eq. (\ref{not ok}),
now the state described by Eq. (\ref{ok}) is not orthogonal to $\left\vert
ok\right\rangle _{L}$ any more. Thus $\bar{F}$ cannot infer that $W$ will
necessarily observe $w=fail$ at time $n:31$ as she did in the above point
(i). That is, the above \textit{Statement} $\bar{F}^{n:02}$ no longer holds,
which also breaks down the logical link between \textit{Statements} $%
F^{n:12} $\ and $F^{n:13}$.

Similarly, unlike Eq. (\ref{pure}), now the state described by Eq. (\ref{mix}%
) is not orthogonal to $\left\vert \overline{ok}\right\rangle _{\bar{L}%
}\otimes \left\vert \downarrow \right\rangle _{S}$. Then $\bar{W}$\ can no
longer be certain\ that there will always be $(\bar{w},z)\neq (\overline{ok}%
,-1/2)$ as he did in the above point (iii). Instead, when $\bar{w}=\overline{%
ok}$\ occurs, both the outcomes $z=+1/2$ and $z=-1/2$\ are possible. That
is, knowing $\bar{w}=\overline{ok}$\ can no longer assure $\bar{W}$\ that
there must be $z=+1/2$. Consequently, the \textit{Statement} $\bar{W}^{n:22}$
does not hold, so that \textit{Statement} $\bar{W}^{n:23}$ cannot be
inferred either.

Thus we can see that the reasoning in page 4 of \cite{qi1571} is no longer
valid, so that the statements in the last two columns of Table 3 of \cite%
{qi1571} cannot be inferred either.

In brief, if we take the point of view that measurements will make
wavefunctions collapse irreversibly so that the information on the phase
difference is erased, then the key reason why the Gedankenexperiment in \cite%
{qi1571} had led to the inconsistent result is that the agents in the
Gedankenexperiment mistakenly described the states of the labs as pure
states (i.e., Eqs. (\ref{not ok}) and (\ref{pure})). But in fact they should
be mixtures (Eqs. (\ref{ok}) and (\ref{mix})). Therefore, when using
Copenhagen interpretation and similar quantum interpretations,
quantum theory will not cause any inconsistent in this Gedankenexperiment so
that Assumption (C) of \cite{qi1571} will not be violated.

It is also worth noting that some versions of the Copenhagen interpretation
does not even allow the Frauchiger-Renner Gedankenexperiment in the first
place. This is because these theories suggested that a quantum state belongs
always to a microscopic object in connection with a measurement apparatus,
and the latter should be described by classical physics, while the
Frauchiger-Renner Gedankenexperiment treated it the other way.
Nevertheless, even if we do not take this point of view, our above analysis
showed that the collapse feature alone is already sufficient for disproving
Frauchiger-Renner's result.

\bigskip

\textit{(2) In quantum interpretation theories where there is no collapse of
wavefunctions.}

Some quantum interpretation theories (e.g., the many-worlds interpretation)
exclude the idea of collapse of wavefunctions. Even some versions of the
Copenhagen interpretation take this viewpoint too. In this case Eq. (\ref%
{pure}) can be taken as the correct description. This is because agent $\bar{%
W}$ can consider $\bar{F}$ and her particles as a unitarily-interacting
closed system, and thus conclude that the post-measurement status is a
coherent superposition. More detailedly, this is done as follows. Let $%
\left\vert init\right\rangle _{\bar{F}}\otimes \left\vert init\right\rangle
_{S}$\ denote the initial state of agent $\bar{F}$ and particle $S$\ right
before she measures\ system $R$\ at time $n:00$, and $\left\vert \bar{h}%
\right\rangle _{\bar{F}}\otimes \left\vert \downarrow \right\rangle _{S}$\
(or $\left\vert \bar{t}\right\rangle _{\bar{F}}\otimes \left\vert
\rightarrow \right\rangle _{S}$) denote the final state that $\bar{F}$
obtains the output $r=heads$\ (or $r=tails$) and sets the spin of $S$ to $%
\left\vert \downarrow \right\rangle _{S}$ (or $\left\vert \rightarrow
\right\rangle _{S}$). From $\bar{W}$'s point of view, instead of applying
the projective operators $\{\left\vert heads\right\rangle _{R}\left\langle
heads\right\vert ,\left\vert tails\right\rangle _{R}\left\langle
tails\right\vert \}$\ on system $R$\ (whose state is in Eq. (\ref{init}))
alone, at time $n:00$ lab $\bar{L}=R\otimes \bar{F}$ and particle $S$\
actually went through a unitary transformation%
\begin{equation}
U_{\bar{L}S}^{00\rightarrow 10}\equiv \left\vert heads\right\rangle
_{R}\left\langle heads\right\vert \otimes U_{\bar{F}S}^{\bar{h}}+\left\vert
tails\right\rangle _{R}\left\langle tails\right\vert \otimes U_{\bar{F}S}^{%
\bar{t}},
\end{equation}%
where $U_{\bar{F}S}^{\bar{h}}$ (or $U_{\bar{F}S}^{\bar{t}}$) is a unitary
transformation on $\bar{F}\otimes S$ that maps $\left\vert init\right\rangle
_{\bar{F}}\otimes \left\vert init\right\rangle _{S}$\ into $\left\vert \bar{h%
}\right\rangle _{\bar{F}}\otimes \left\vert \downarrow \right\rangle _{S}$\
(or $\left\vert \bar{t}\right\rangle _{\bar{F}}\otimes \left\vert
\rightarrow \right\rangle _{S}$). Then the resultant state of $\bar{L}%
\otimes S$ is%
\begin{eqnarray}
\left\vert 10\right\rangle _{\bar{L}S} &\equiv &U_{\bar{L}S}^{00\rightarrow
10}(\left\vert init\right\rangle _{R}\otimes \left\vert init\right\rangle _{%
\bar{F}}\otimes \left\vert init\right\rangle _{S})  \nonumber \\
&=&\sqrt{\frac{1}{3}}\left\vert heads\right\rangle _{R}\otimes \left\vert
\bar{h}\right\rangle _{\bar{F}}\otimes \left\vert \downarrow \right\rangle
_{S}  \nonumber \\
&&+\sqrt{\frac{2}{3}}\left\vert tails\right\rangle _{R}\otimes \left\vert
\bar{t}\right\rangle _{\bar{F}}\otimes \left\vert \rightarrow \right\rangle
_{S}.  \label{pure2}
\end{eqnarray}%
Denoting%
\begin{equation}
\left\vert \bar{h}\right\rangle _{\bar{L}}=\left\vert heads\right\rangle
_{R}\otimes \left\vert \bar{h}\right\rangle _{\bar{F}}
\end{equation}%
and%
\begin{equation}
\left\vert \bar{t}\right\rangle _{\bar{L}}=\left\vert tails\right\rangle
_{R}\otimes \left\vert \bar{t}\right\rangle _{\bar{F}},
\end{equation}%
we successfully obtain Eq. (\ref{pure}) from Eq. (\ref{pure2}).

From this deduction process, it can be seen that if we take Eq. (\ref{pure})
as valid, we actually accepted the following logic:

\textit{For any agent (such as }$\bar{W}$\textit{) who knows nothing about
what happened inside the lab (e.g., which outcome }$\bar{F}$\textit{\ had
obtained), the state of the lab should be described as a quantum
superposition, in which all terms corresponding to all possible outcomes
inside the lab are included.}

But in this scenario, another problem will arise in the original reasoning
in \cite{qi1571}. As stressed in \cite{qi1571}, all agents employ the same
theory throughout the Gedankenexperiment. Therefore, when agent $F$\
measures $S$ at time $n:10$, from other agents' point of view, lab $%
L=S\otimes F$\ should also be considered as a unitarily-interacting system.
Let $\left\vert init\right\rangle _{F}$\ denote the initial state of agent $%
F $. Then the unitary transformation corresponding to this process can be
expressed as%
\begin{equation}
U_{L}^{10\rightarrow 20}\equiv \left\vert \downarrow \right\rangle
_{S}\left\langle \downarrow \right\vert \otimes U_{F}^{(-)}+\left\vert
\uparrow \right\rangle _{S}\left\langle \uparrow \right\vert \otimes
U_{F}^{(+)},  \label{U10}
\end{equation}%
where $U_{F}^{(-)}$ (or $U_{F}^{(+)}$) is a unitary transformation on $F$
that maps $\left\vert init\right\rangle _{F}$\ into $\left\vert
-1/2\right\rangle _{F}$\ (or $\left\vert +1/2\right\rangle _{F}$). Here $%
\left\vert -1/2\right\rangle _{F}$\ (or $\left\vert +1/2\right\rangle _{F}$)
denotes the state of $F$\ when the state of lab $L$\ is $\left\vert
-1/2\right\rangle _{L}$\ (or $\left\vert +1/2\right\rangle _{L}$), i.e.,%
\begin{equation}
\left\vert -\frac{1}{2}\right\rangle _{L}=\left\vert \downarrow
\right\rangle _{S}\otimes \left\vert -\frac{1}{2}\right\rangle _{F}
\end{equation}%
and%
\begin{equation}
\left\vert +\frac{1}{2}\right\rangle _{L}=\left\vert \uparrow \right\rangle
_{S}\otimes \left\vert +\frac{1}{2}\right\rangle _{F}.
\end{equation}%
Applying Eq. (\ref{U10}) on Eq. (\ref{pure2}) (i.e., Eq. (\ref{pure})), we
know that the resultant state of $\bar{L}\otimes L=\bar{L}\otimes S\otimes F$
is%
\begin{eqnarray}
&&I_{\bar{L}}\otimes U_{L}^{10\rightarrow 20}(\left\vert 10\right\rangle _{%
\bar{L}S}\otimes \left\vert init\right\rangle _{F})  \nonumber \\
&=&\sqrt{\frac{1}{3}}\left\vert \bar{h}\right\rangle _{\bar{L}}\otimes
\left\vert \downarrow \right\rangle _{S}\otimes \left\vert -\frac{1}{2}%
\right\rangle _{F}  \nonumber \\
&&+\sqrt{\frac{2}{3}}\left\vert \bar{t}\right\rangle _{\bar{L}}\otimes \sqrt{%
\frac{1}{2}}\left( \left\vert \downarrow \right\rangle _{S}\otimes
\left\vert -\frac{1}{2}\right\rangle _{F}+\left\vert \uparrow \right\rangle
_{S}\otimes \left\vert +\frac{1}{2}\right\rangle _{F}\right)  \nonumber \\
&=&\sqrt{\frac{1}{3}}\left\vert \bar{h}\right\rangle _{\bar{L}}\otimes
\left\vert -\frac{1}{2}\right\rangle _{L}  \nonumber \\
&&+\sqrt{\frac{2}{3}}\left\vert \bar{t}\right\rangle _{\bar{L}}\otimes \sqrt{%
\frac{1}{2}}\left( \left\vert -\frac{1}{2}\right\rangle _{L}+\left\vert +%
\frac{1}{2}\right\rangle _{L}\right)  \nonumber \\
&=&\sqrt{\frac{2}{3}}\sqrt{\frac{1}{2}}\left( \left\vert \bar{h}%
\right\rangle _{\bar{L}}+\left\vert \bar{t}\right\rangle _{\bar{L}}\right)
\otimes \left( \left\vert -\frac{1}{2}\right\rangle _{L}+\frac{1}{2}%
\left\vert +\frac{1}{2}\right\rangle _{L}\right)  \nonumber \\
&&+\sqrt{\frac{1}{12}}\left\vert \overline{ok}\right\rangle _{\bar{L}%
}\otimes \left( \left\vert ok\right\rangle _{L}-\sqrt{\frac{1}{2}}\left(
\left\vert -\frac{1}{2}\right\rangle _{L}+\left\vert +\frac{1}{2}%
\right\rangle _{L}\right) \right) ,  \nonumber \\
&&  \label{final}
\end{eqnarray}
where $I_{\bar{L}}$\ is the identity operator on lab $\bar{L}$. This
equation shows that the outcome $(\bar{w},w)=(\overline{ok},ok)$ can occur
with probability $1/12$, which is in agreement with Eq. (7) of \cite{qi1571}%
. But the important point is: since all agents employ the same theory, not
only $W$, but also $\bar{F}$\ and $\bar{W}$\ are fully aware of this
equation too. Note that Eq. (\ref{final}) contains not only the term $\sqrt{%
1/2}\left( \left\vert -1/2\right\rangle _{L}+\left\vert +1/2\right\rangle
_{L}\right) $, but also other terms such as $\left\vert ok\right\rangle _{L}$%
. Thus it brings forth the following question: when $W$ is going to measure
lab $L$ at time $n:30$, which one is the correct description of the state of
$L$, Eq. (\ref{final}) or Eq. (\ref{not ok})?

To answer the question, we must note that different agents will have
different point of view. Remind that $\bar{W}$ and\ $W$\ both take the
perspective that no collapse occurs in measurements (otherwise our above
point (1) applies), and they do not know whether $\bar{F}$ got $r=tails$ or
not before performing their own measurements. Therefore, following the same
logic stated above that made Eq. (\ref{pure}) valid, from $\bar{W}$'s\ and $%
W $'s point of view, all terms corresponding to both $\left\vert \bar{h}%
\right\rangle _{\bar{L}}$ and $\left\vert \bar{t}\right\rangle _{\bar{L}}$
should remain. That is, $\bar{W}$ and\ $W$\ will take Eq. (\ref{final}) as
correct, so they will not be surprised to find that the outcome $\bar{w}=%
\overline{ok}$ and $w=ok$ can occur in the same round.

But from $\bar{F}$'s point of view, she knows that she got $r=tails$. Thus
it seems that she herself could take Eq. (\ref{not ok}) as the correct
description. 
However, when she is to infer the result of other agents, 
it will be a different story. As she knows that \textquotedblleft $W$\textit{%
\ does not know whether I got }$r=tails$\textit{\ or not}\textquotedblright\
and they all use the same quantum theory, if she\ is wise enough, she should
be able to infer that \textquotedblleft $W$\textit{\ will take Eq. (\ref%
{final}) instead of Eq. (\ref{not ok}) as the correct description of the
state of }$L$\textit{\ before performing his measurement}\textquotedblright
. As a result, $\bar{F}$ should no longer make \textit{Statement} $\bar{F}%
^{n:02}$ that \textquotedblleft I am certain that $W$ will observe $w=fail$
at time $n:31$.\textquotedblright . Instead, she should infer from Eq. (\ref%
{final}) that

\textit{Statement} $\bar{F}^{n:02\ast }$: \textquotedblleft I am certain
that when $\bar{W}$ find $\bar{w}=\overline{ok}$\ at time $n:21$, $W$ will
have a nonzero probability to observe $w=ok$ at time $n:31$%
.\textquotedblright

Consequently, all those statements (especially \textit{Statements} $F^{n:13}$
and $\bar{W}^{n:23}$) in Table 3 of \cite{qi1571} that were inferred from
the old \textit{Statement} $\bar{F}^{n:02}$ will no longer hold. Also, when
agent $\bar{W}$ find $\bar{w}=\overline{ok}$, he can make his own judgement
from Eq. (\ref{final}) directly that

\textit{Statement} $\bar{W}^{n:23\ast }$: \textquotedblleft I am certain
that $W$ will have a nonzero probability to observe $w=ok$ at time $n:31$%
.\textquotedblright

Therefore, the inconsistent results will no longer exist.

In other words, if $\bar{F}$ insists that both Eqs. (\ref{not ok}) and (\ref%
{pure}) are the correct description of the states of the corresponding
systems even from $W$'s\ point of view, then she is in fact not using the
same quantum theory consistently throughout the whole Gedankenexperiment.
That is, upon writing the state of $\bar{L}\otimes S$ as Eq. (\ref{pure}),
she is taking the perspective that no collapse of wavefunctions occurs in
measurement, so that both the terms corresponding to $\left\vert \downarrow
\right\rangle _{S}$ and $\left\vert \rightarrow \right\rangle _{S}$\ exist.
But when she think that Eq. (\ref{not ok}) is the correct description of the
state of lab $L$ even from $W$'s\ point of view (so that she can infer
\textit{Statement} $\bar{F}^{n:02}$), she actually take collapse into her
picture because only the term corresponding to $\left\vert \rightarrow
\right\rangle _{S}$\ remains in Eq. (\ref{not ok}) while the term
corresponding to $\left\vert \downarrow \right\rangle _{S}$ disappeared.
This is how the inconsistent results were introduced.

\section{The role of interpretation theories}

Taking points (1) and (2) in the previous section together, we see that no
matter we take an interpretation theory with or without collapse, the
contradiction result in the Frauchiger-Renner Gedankenexperiment will always
be avoided.

But this situation by no means indicates that interpretation theory is not
important to quantum mechanics. On the contrary, it highlights the need for
a researcher to make a specific choice on the interpretation theory, and use
it consistently throughout the analysis of quantum phenomena. Otherwise, if
we apply the quantum mechanical formalism without having a clear picture on
the interpretation, there will be the chance to ask an improper question,
which bonds to have no correct answer.

In the current case, we see the importance to always have in mind a fixed
picture of what happens during quantum measurement, even though it does not
matter whether this picture allows collapse or not. In some interpretation
theories (e.g., Everett's many-worlds interpretation), the quantum
mechanical formalism comprises states and unitary evolution only \cite%
{qi1813}. The Born rule is not included as an axiom. Instead, it is taken as
a part of the interpretation and has to be derived independently, so that
the very meaning of the phrase \textquotedblleft outcome of a
measurement\textquotedblright\ can be defined. In fact, Born's original idea
was to apply a quantum state only to ensembles. But Assumption (Q) of \cite%
{qi1571} seems to have made the tacit premise that an individual system is
associated with a certain quantum state. Even so, if we stick to a single
interpretation, either with or without collapse, and apply it consistently
to analysis the statements of the agents, then there will be no problem, as
it was shown in the previous section. On the contrary, Ref. \cite{qi1571} made an
inconsistent logic that Eq. (\ref{pure}) was derived
without taking collapse into consideration, while \textit{Statement} $\bar{F}%
^{n:02}$ seems to miss the fact that if quantum states always evolve
unitarily without collapse, then the outcome of $w$ should remain in a
superposition instead of taking one of the value with certainty, and Eq.  (\ref{not ok}) is actually half-way between the collapse picture (where it should be replaced by Eq.(\ref{ok})) and the non-collapse picture (where it should be replaced by Eq.(\ref{final})). That is,
the shifting use between collapse and non-collapse interpretations in \cite%
{qi1571} is the cause that eventually led to the inconsistency among \textit{Statement} $\bar{F}^{n:02}$, Eq. (\ref{not ok}) and Eq. (%
\ref{pure}).

\section{Conclusions and comparison with related works}
\begin{table*}
\caption{Comparison between our results and these of Refs. \cite{qi1631,qi1572,qi1575,qi1592,qi1608,qi1680,qi1683}.}
\label{tab:1}       
\begin{tabular} {p{30pt}p{280pt}}
\hline\noalign{\smallskip}
Ref. no. & Additional hypothese or approach for making quantum theory consistent in Frauchiger-Renner Gedankenexperiment\\
\noalign{\smallskip}\hline\noalign{\smallskip}
\cite{qi1631} & Proposed a no-go theorem for observer-independent facts, which implies that the states referring to outcomes of different
observers in such a Gedankenexperiment cannot be defined without referring
to the specific experimental arrangements of the observers. \\
\cite{qi1572} & None.* \\
\cite{qi1575} & Introduced a
different description of quantum measurements (i.e.,
Hypothesis 2 in \cite{qi1575}). \\
\cite{qi1592} & Added quantum memories a la
Deutsch \cite{Deutsch} to the analysis of the Gedankenexperiment. \\
\cite{qi1608} & Used relational
formulation of quantum measurement \cite{RQM} in the analysis. \\
\cite{qi1680} & None.* \\
\cite{qi1683} & Used random measurement process to prove that assumption (Q) is too broad and meaningless. \\
Ours & None. \\
\noalign{\smallskip}\hline
  & * See the last paragraph of Sect. V for details.
\end{tabular}
\end{table*}

In summary, none of the quantum interpretation theories will violate
Assumption (C) when being applied correctly. On one hand, if we take the
perspective of point (1) in Sect. III that wavefunctions will
collapse when being measured, then Eqs. (\ref{not ok}) and (\ref{pure}) are
incorrect and should be replaced by Eqs. (\ref{ok}) and (\ref{mix}). On the
other hand, if we take the perspective of point (2) that all measurements
can be phrased as a unitarily-interacting process without involving collapse
of wavefunctions, then Eq. (\ref{pure}) could remain
valid but \textit{Statement} $\bar{F}^{n:02}$ becomes incorrect. Either way,
it leads to \textit{Conclusion 1} of our current paper:

\textit{The inconsistent results in \cite{qi1571} come from its own faulty
reasoning, and not from quantum theory itself. The fault is that they had
not applied a single specific interpretation theory consistently throughout
the analysis of their Gedankenexperiment.}

\bigskip

Moreover, Table 4 of \cite{qi1571} already showed that the Copenhagen
interpretation can satisfy both assumptions (Q) and (S). Therefore,
combining with our result, we reach \textit{Conclusion 2}:

\textit{The Copenhagen interpretation can fulfil all the three assumptions
(C), (Q), and (S) simultaneously without causing contradictory statements,
thus disprove Theorem 1 of \cite{qi1571}.}

\bigskip

Around the same time we completed the initial version of the current paper
\cite{Hearxiv}, we came to aware that other researchers also proposed
different solutions to the inconsistency. Table 1 listed the main claims of these works (showed in chronological order).


From the table we can see that our result has a significant difference from
these of Refs. \cite{qi1631,qi1575,qi1592,qi1608,qi1683}. That
is, to avoid the inconsistency in the Gedankenexperiment,
they all suggested to introduce additional hypothese or approach to quantum theory, while our conclusion 1 showed that there is no such need.
Refs. \cite{qi1572,qi1680} made no modification to quantum theory either. But Ref. \cite{qi1572} actually modified the Gedankenexperiment, so that it is not dealing with the original problem in \cite{qi1571}. In Ref. \cite{qi1680} the author stated that \textquotedblleft I find it hard to see how this assumption (assumption (C)) is violated in any interpretation of quantum mechanics, except possibly QBism\textquotedblright . But it has not pointed out rigorously which equations or statements of \cite{qi1571} went wrong when not using the Consistent Histories interpretation, and what is the key reason that led to this mistake. On the contrary, we revealed that all interpretation theories can satisfy Assumption (C) without modification, and Ref. \cite{qi1571} thought it cannot be satisfied simply because they themselves had used the interpretation theories inconsistently, so that they mistakenly believed that \textit{Statement} $\bar{F}^{n:02}$, Eq. (\ref{not ok}) and Eq. (\ref{pure}) can be valid simultaneously.


\section*{Acknowledgements}

We thank the anonymous reviewer for sharing the ideas that leads to Sect.
IV. The work was supported in part by Guangdong Basic and Applied Basic Research
Foundation under grant No. 2019A1515011048.





\end{document}